\documentclass[aps,twocolumn,nofootinbib]{revtex4}

\usepackage{graphicx}
\usepackage{dcolumn}
\usepackage{bm}

\usepackage{amssymb,amsfonts}
\usepackage{epsfig}
\usepackage{epstopdf}
\usepackage[all]{xy}
\usepackage{amsthm}
\usepackage{dcolumn}
\usepackage{hyperref}
\usepackage{url}
\usepackage{textcomp}
\usepackage{csquotes}
\usepackage{dsfont}

\newcommand{\be}{\begin{equation}}
\newcommand{\ee}{\end{equation}}
\newcommand{\bea}{\begin{eqnarray}}
\newcommand{\nn}{\nonumber}
\newcommand{\eea}{\end{eqnarray}}

\def\inbar{\,\vrule height1.5ex width.4pt depth0pt}
\def\IR{\relax{\rm I\kern-.18em R}}
\def\IC{\relax\hbox{$\inbar\kern-.3em{\rm C}$}}

\begin{document}

\title{Covariant and infrared-free graviton two-point function in de Sitter spacetime II}

\author{Hamed Pejhan\footnote{h.pejhan@piau.ac.ir}} \author{Surena Rahbardehghan\footnote{s.rahbardehghan@iauctb.ac.ir}}
\affiliation{Department of Physics, Science and Research Branch, Islamic Azad University, Tehran, Iran}

\begin{abstract}
The solution to the linearized Einstein equation in de Sitter (dS) spacetime and the corresponding two-point function are explicitly written down in a gauge with two parameters `$a$' and `$b$'. The quantization procedure, independent of the choice of the coordinate system, is based on a rigorous group theoretical approach. Our result takes the form of a universal spin-two (transverse-traceless) sector and a gauge-dependent spin-zero (pure-trace) sector. Scalar equations are derived for the structure functions of each part. We show that the spin-two sector can be written as the resulting action of a second-order differential operator (the spin-two projector) on a massless minimally coupled scalar field (the spin-two structure function). The operator plays the role of a symmetric rank-$2$ polarization tensor and has a spacetime dependence. The calculated spin-two projector grows logarithmically with distance and also no dS-invariant solution for either structure functions exist. We show that the logarithmically growing part and the dS-breaking contribution to the spin-zero part can be dropped out, respectively, for suitable choices of parameters `$a$' and `$b$'. Considering the transverse-traceless graviton two-point function, however, shows that dS breaking is universal (cannot be gauged away). More exactly, if one wants to respect the covariance and positiveness conditions, the quantization of the dS graviton field (as for any gauge field) cannot be carried out directly in a Hilbert space and involves unphysical negative norm states. However, a suitable adaptation (Krein spaces) of the Gupta-Bleuler scheme for massless fields, based on the group theoretical approach, enables us to obtain the corresponding two-point function satisfying the conditions of locality, covariance, transversality, index symmetrizer, and tracelessness.
\end{abstract}
\maketitle

\section{Introduction}
\label{sec:intro}
The relevance of de Sitter spacetime (for a review see e.g. \cite{SHawking1973}) to certain cosmological models such as inflationary epoch during the early moments of the Universe \cite{GSmoot396,GHinshaw288,AGuth347,ALinde389,AAlbrecht1220} has brought increasing attention to quantum field theory on this background in recent years. Moreover, cosmological observations show that the expansion of our Universe is accelerating, so that, it might evolve into a de Sitter stage in the future \cite{ARiess1009}. Another interest in dS space stems from the fact that it is the maximally symmetric solution of Einstein equation with positive cosmological constant. Therefore, the study of the linear perturbations of Einstein gravity around the de Sitter metric (the dS linear gravity) and the associated graviton two-point functions, which represent correlation of vacuum fluctuation in the gravitational field, are of particular importance in dS space.

Investigation of graviton two-point function in de Sitter spacetime has been performed extensively in the literature from various point of views \cite{BAllen3670,BAllen743,BAllen813,EFloratos373,IAntoniadis1037,NTsamis217,IAntoniadis437}. One of the main subjects in analyzing graviton propagator, more exactly, the question that whether infrared (IR) divergences are restricted to the gauge sector of linearized gravity or they also appear in the physical sector, has been the origin of controversy for over three decades. Many authors have studied the subject and utilized IR divergences associated with the graviton propagator, which explicitly break de Sitter invariance, to obtain physical results, e.g., instability in de Sitter space, decrease of cosmological constant in time,\footnote{For a criticism about some obtained results and refutation see for instance \cite{JGarriga024021,NTsamis028501}.} inflationary cosmology, and change in some coupling constants \cite{IAntoniadis1319,LFord710,NTsamis292,NTsamis351,NTsamis1,
NTsamis105006,SGiddings023,SGiddings063528,SGiddings083538,
HKitamoto124007,Hkitamoto124004,RWoodard1430020,SMiao122301,EKahya022304,PMora122502,SMiao104004}. On the other side and quite contrary to the above point of view, many believe that IR-divergent part of the graviton two-point function is gauge artifact and hence is not physical. To review the viewpoint for various gauges and coordinate systems, one can refer to \cite{LFord1601,SHawking063502,AHiguchi3077,AHiguchi2933,AHiguchi397,AHiguchi245012,
AHiguchi3005,RBernar024045,AHiguchi4317,AHiguchi124006,MFaizal124021,IMorrison13021860}.

In this paper, neglecting the graviton-graviton interactions, we proceed with the examination of the graviton two-point function in de Sitter spacetime through a group theoretical approach. More precisely, we generalize our previous work \cite{HPejhan044016} and write the linearized Einstein equation in a gauge with two parameters `$a$' and `$b$'. This procedure not only allows us to handle the logarithmic divergences of the spin-two projector and the dS-breaking contribution to the spin-zero part of the graviton two-point function, but also provides the opportunity to suppress the mathematical shortcomings in the guage-fixing procedure of our previous work (for a detailed discussion about these shortcomings, refer to Sec. III-B, Part 1. Comment on the gauge-fixing procedure in \cite{HPejhan044016}). Moreover, we remarkably show that the IR divergences associated with the transverse-traceless graviton two-point function are completely independent of the choice of the gauge-fixing parameters.

At the beginning, in Sec. II, we start from the linearized Einstein equation given in de Sitter intrinsic coordinates and rewrite it by using the ambient space formalism in terms of the coordinate-independent de Sitter-Casimir operators. With respect to the spectral values of these Casimir operators, the unitary irreducible representations (UIRs) of the dS group are classified \cite{Dixmier9,Takahashi289}. Now, it is our goal to find an expression for the (linear) gravity, the field solution, in terms of covariant derivative projection operators acting on some scalar functions.

The transverse-traceless part of the field (${\cal{K}}_{\alpha\beta}^{TT}(x)$) is considered in Sec. III-A and B (the pure-trace sector will be investigated in Sec. IV). In this regard, first, the group theoretical content of the field equation is studied. Then, the source of the mathematical shortcomings of the previous work are explicitly discussed. More exactly, we address that how the mistake was entered and more importantly how by utilizing a proper gauge-fixing procedure the main achievements of the previous work is preserved. Indeed, in complete agreement with the context of the de Sitter group theory \cite{Gazeau2533,Gazeau507}, and previous works (see for instance \cite{AHiguchi4317,AHiguchi124006}), we show that the value ${a}=5/3$ corresponds to the minimal (or \enquote{optimal}) choice, without any logarithmic singularity. Choosing ${a}=5/3$, we construct the transverse-traceless part of the field solution in terms of the massless minimally coupled (MMC) scalar field $\phi_m$ and a projection operator
$${\cal{K}}^{TT}(x) = {\cal{D}}^{TT}(x,\partial)\phi_m.$$

The solutions, however, do not constitute a closed set under the action of the dS group for any gauge field.\footnote{Here, we especially emphasize the covariance aspect which should be understood in the sense of the action of the dS group.} Actually, an explicit computation gives that the problematic element of the solution is quite relevant to its structure function (the MMC scalar field). To present a deeper insight into the problem, we notice that comprehensive studies of the quantization of the MMC scalar field have shown that (see for instance \cite{Gazeau+yo} and references therein) in obtaining a covariant construction of the propagator function for the field on the Euclidean continuation of dS space, $S^4$, one confronts the obstacle that the Laplace-Beltrami operator $\square_H$ has a normalizable zero-frequency mode (more accurately a constant mode). Therefore, no dS-invariant propagator inverse for the wave operator $\square_H$ exists; the infrared divergence appears.\footnote{It is worth mentioning that in some coordinates such as the Poincar\'{e} patch, which is the part of de Sitter space pertinent to inflation, there is no constant mode at all. Indeed, the nonconvergence of the Fourier integral is responsible for the infrared problems, which must be cut off at some low momentum to obtain a finite result.} It should be emphasized that this result is not an artifact of the Euclidean continuation since Allen has proved that \cite{BAllen3136,BAllen3771} no de Sitter covariant Fock vacuum for the MMC scalar field exists. Note that, the norm of the so-called zero mode is positive, nonetheless, it is not part of the Hilbertian structure of the one-particle sector \cite{de Bievre6230,Gazeau1415,Garidi,Bertola}. More clearly, the action of the dS group on it generates all the negative frequency solutions (with regard to the conformal time) of the field equation. On the other hand, another difficulty appears when dealing with fields involving a gauge invariance (the de Sitter MMC free field Lagrangian ${\cal{L}}=\sqrt{|g|}\partial_\mu \phi_m \partial^\mu \phi_m$ is invariant under a gaugelike global transformation $\phi_m \longrightarrow \phi_m + \mbox{`constant'}$). The Gupta-Bleuler formalism was invented to handle both covariance and gauge invariance in quantum electrodynamics. Therefore, it is not surprising that an similar construction performes the same task for the MMC scalar field on dS spacetime. On this basis and in consistency with Allen's theorem \cite{BAllen3136,BAllen3771}, it has been shown that \cite{Gazeau1415,de Bievre6230,Garidi,Bertola} the dS breaking and the associated infrared divergence of the MMC scalar field disappear if one uses the Gupta-Bleuler type vacuum (the Krein-Gupta-Bleuler (KGB) vacuum) defined by de Bi\`{e}vre, Renaud \cite{de Bievre6230} and Gazeau, Renaud, Takook \cite{Gazeau1415}.

As noted above, the transverse-traceless linearized gravitons suffer from the same difficulty as the MMC scalar field. Therefore, to construct the fully de Sitter covariant and infrared-free two-point function for the transverse-traceless linearized gravitons, the KGB quantization should be taken into account (see Sec. III-C); the minimal space on which the Fock space should be constructed to preserve the covariance of the full theory under the full dS group $SO_o(1, 4)$\footnote{The subscript $0$ refers to the subgroup of $SO(1,4)$ connected to the identity.} is the Krein spcae (the direct sum of Hilbert and anti-Hilbert space). Here, it must be emphasized that including the negative norm states in the theory, which is the price to pay in order to obtain a fully covariant theory, endangers the analyticity of the final result for the graviton two-point function. Indeed, respecting the above statements, the only fully dS-covariant two-point function for linearized gravitons which naturally appears is the commutator that is not of the positive type and it does not allow us to select physical states (for a detailed discussion see \cite{HPejhan044016}). Again, the crucial point is that any definition \emph{a priori} of different two-point functions, like Wightman or Hadamard functions, cannot yield a covariant theory; there exists no nontrivial covariant two-point function of the positive type for the MMC scalar field on dS space \cite{BAllen3136,BAllen3771,de Bievre6230,Gazeau1415,Garidi,Bertola}. In spite of the presence of negative norm modes in the theory, however, it must be underlined that no negative energy can be measured: expressions as $\langle n_{k_1} n_{k_2} ...|T_{00}|n_{k_1} n_{k_2} ... \rangle$ are always positive \cite{Gazeau1415,de Bievre6230}.

Interestingly, our group theoretical approach to the dS linearized gravitons supports the results expressed by Woodard \emph{et al.}: \enquote{\emph{one encounters contrary statements in the mathematical physics literature, so that the IR divergence of graviton propagator is gauge artifact and hence can be gauged away (e.g. see \cite{MFaizal124021} and references therein), but close examination reveals that the authors admit they are constructing a formal solution to the propagator equation which is not a true propagator\footnote{\emph{Reminder:} the only fully dS-covariant two-point function for linearized gravitons which naturally appears is the commutator that is not of the positive type ... .} by the illegitimate technique of adding negative norm states to the theory}} \cite{RWoodard1430020}. \enquote{\emph{Indeed, including negative norm states in the mode sum is the only way to avoid de Sitter breaking and also the problematic nature of analytical continuations}} \cite{SMiao104004}.

At the end, the pure-trace sector of the theory is investigated in Sec. IV. We show that the pathological large distance behavior associated with this part is gauge-dependent and, hence, a proper gauge-fixing procedure can eliminate IR divergences and preserve dS invariance. We also show that there is a suitable value for parameter `$b$' for which the spin-zero sector is written in terms of the dS massless conformally coupled scalar field.

Finally, a brief conclusion is given in Sec. V.

\section{The field equation}
\label{sec:equiv}
In this section, by splitting our metric into a dS fixed background $\hat{g}_{\mu\nu}$ and a small fluctuation $h_{\mu\nu}$; $g_{\mu\nu}=\hat{g}_{\mu\nu}+h_{\mu\nu}$, we investigate linear perturbations of Einstein gravity around the dS metric. Pursuing this path, the following gauge invariance is the translation of the reparametrization invariance
\begin{equation}\label{2.4} h_{\mu\nu}\rightarrow h_{\mu\nu} + (\nabla_{\mu}\Xi_{\nu} + \nabla_{\nu}\Xi_{\mu}),\end{equation}
$\Xi_{\nu}$ is an arbitrary vector field. The wave equation for massless\footnote{Note that, thanks to the maximal symmetry of (anti-)de Sitter spaces, the mass concept can be defined precisely on these spacetimes \cite{Gazeau304008,Flato415}.} tensor fields $h_{\mu\nu}$ propagating on dS background therefore would be
\begin{eqnarray} \label{2.3}
(\Box_H + 2H^2)h_{\mu\nu} - (\Box_H - H^2)\hat{g}_{\mu\nu}h' - 2\nabla_{(\mu}\nabla^\rho h_{\nu)\rho} \nonumber \\ + \hat{g}_{\mu\nu}\nabla^\lambda\nabla^\rho h_{\lambda\rho} + \nabla_\mu\nabla_\nu h'=0,
\end{eqnarray}
$H$ stands for the Hubble constant, $\nabla^\nu$ for the dS covariant derivative, $\Box_H= \hat{g}_{\mu\nu}\nabla^\mu\nabla^\nu$ for the Laplace-Beltrami operator and $h'=\hat{g}^{\mu\nu}h_{\mu\nu}$. In what follows, a generalization of the Lorentz/harmonic gauge condition will be considered as the gauge-fixing term
\begin{equation}\label{2.5}
\nabla^\mu h_{\mu\nu} = {b}\nabla_\nu h',
\end{equation}
where `$b$' is an arbitrary constant.

Here, with respect to the Wigner theorem and in analogy with the Minkowskian cases, we wish to construct the dS tensor field equation (\ref{2.3}) as an eigenvalue equation of a de Sitter group Casimir operator. In this regard, it will be convenient to employ the ambient space coordinates that makes manifest the $SO_0(1,4)$ invariance; the dS spacetime is described as a one-sheeted hyperboloid embedded in a 5-dimensional Minkowski spacetime ($\alpha,\beta=0,1,2,3,4$)
\begin{eqnarray}\label{2.1}
{X_{H}}  = \{ x \in {\Re}^5 ; x^2={\eta}_{\alpha\beta} {x^{\alpha}} {x^\beta} =  -H^{-2}\},
\end{eqnarray}
where $\eta_{\alpha\beta}=$ diag$(1,-1,-1,-1,-1)$. The dS metric then would be the induced metric on the dS hyperboloid
\begin{equation}\label{2.2}
ds^2=\eta_{\alpha\beta}dx^{\alpha}dx^{\beta}|_{x^2=-H^{-2}}=\hat{g}_{\mu\nu}dX^{\mu}dX^{\nu},
\end{equation}
where the $X^\mu$'s are intrinsic spacetime coordinates ($\mu,\nu=0,1,2,3$).

In these notations, the tensor field ${\cal{K}}_{\alpha\beta}(x)$ can be viewed as a homogeneous function in the ${\Re}^5$ variables $x^\alpha$ with some arbitrarily chosen degree $\sigma$,
\begin{equation}\label{2.6}
x^\alpha \frac{\partial}{\partial x^\alpha}{\cal{K}}_{\beta\gamma}(x) = x\cdot\partial {\cal{K}}_{\beta\gamma}(x) = \sigma{\cal{K}}_{\beta\gamma}(x),
\end{equation}
while, the transversality condition guarantees that the direction of ${\cal{K}}$ lies in the dS spacetime
\begin{equation}\label{2.7}
x^\alpha {\cal{K}}_{\alpha\beta}(x) = x^\beta {\cal{K}}_{\alpha\beta}(x)\;\Big(\equiv x\cdot {\cal{K}}(x)\Big) =0.
\end{equation}
Respecting the importance of this transversality property of dS fields, defining the symmetric transverse projector $\theta_{\alpha \beta}=\eta_{\alpha \beta}+H^2x_\alpha x_\beta$ enables us to construct transverse entities such as the transverse derivative
\begin{equation}\label{2.9}
\bar{\partial}_\alpha=\theta_{\alpha \beta}\partial^\beta=\partial_\alpha+H^2x_\alpha x\cdot\partial,\;\;\;\; x\cdot\bar{\partial}=0.
\end{equation}
$\theta_{\alpha \beta}$ is actually the transverse form of the de Sitter metric
$$\hat{g}_{\mu\nu}=\frac{\partial x^\alpha}{\partial X^\mu}\frac{\partial x^\beta}{\partial X^\nu}\theta_{\alpha \beta}.$$

On the other hand, considering the above notations, the second order Casimir operator of the dS group can be easily defined in terms of the self-adjoint $L_{\alpha\beta}$ representatives of the Killing vectors\footnote{A familiar realization of the Lie algebra of the dS group is the one generated by the ten Killing vectors $K_{\alpha\beta} = x_{\alpha}\partial_{\beta}-x_{\beta}\partial_{\alpha}$.} \cite{Gazeau2533,Gazeau507}
\begin{equation} \label{2casimir}
Q_2=-\frac{1}{2}L^{\alpha\beta}L_{\alpha\beta}= -\frac{1}{2}(M^{\alpha\beta} + \Sigma^{\alpha\beta})(M_{\alpha\beta} + \Sigma_{\alpha\beta}),
\end{equation}
where the action of the orbital and the spinorial parts are respectively defined by
\begin{eqnarray}
M_{\alpha\beta}\equiv-i(x_{\alpha}\partial_{\beta}-x_{\beta}\partial_{\alpha}),
\end{eqnarray}
\begin{eqnarray}
{\Sigma}_{\alpha\beta}{\cal{K}}_{\gamma\delta}\equiv-i(\eta_{\alpha\gamma}{\cal{K}}_{\beta\delta}-\eta_{\beta\gamma} {\cal{K}}_{\alpha\delta} + \eta_{\alpha\delta}{\cal{K}}_{\gamma\beta}-\eta_{\beta\delta}{\cal{K}}_{\gamma\alpha}),\;\;
\end{eqnarray}
and the subscript $2$ refers to the fact that the carrier space is constituted by second rank tensors. On this basis, the action of $Q_2$ on ${\cal{K}}$ would be
\begin{equation}\label{2.19} Q_2{\cal{K}}=(Q_0-6){\cal{K}}+2\eta{\cal{K}}'+2{\cal{S}}x\partial\cdot{\cal{K}}-2{\cal{S}}\partial x\cdot{\cal{K}}, \end{equation}
in which ${\cal{S}}$ is the symmetrizer operator, ${\cal{S}}\xi_\alpha \omega_\beta=\xi_\alpha \omega_\beta+\xi_\beta \omega_\alpha$, and $Q_0$ is the scalar part of the Casimir operator, $Q_0=-\frac{1}{2}M_{\alpha\beta}M^{\alpha\beta}=-H^{-2}(\bar{\partial})^2$.

Now, considering the above identities and the fact that the \enquote{intrinsic} field $h_{\mu\nu}(X)$ is locally specified by the \enquote{transverse} tensor field ${\cal{K}}_{\alpha\beta}(x)$,
\begin{equation}\label{2.10}
h_{\mu\nu}(X)=\frac{\partial x^\alpha}{\partial X^\mu}\frac{\partial x^\beta}{\partial X^\nu}{\cal{K}}_{\alpha\beta}\big( x(X) \big),
\end{equation}
one can easily present the field equation for ${\cal{K}}_{\alpha\beta}$ in terms of the second order Casimir operator \cite{Dehghani064028,Fronsdal848}
\begin{equation}\label{2.21} (Q_2+6){\cal{K}}(x)+D_2\partial_2\cdot{\cal{K}}(x)=0, \end{equation}
in which
\begin{equation}\label{2.15} D_2K=H^{-2}{\cal{S}}(\bar{\partial}-H^2x)K, \end{equation}
and $\partial_2\cdot$, the generalized divergence on the dS hyperboloid, is
\begin{equation}\label{2.16} \partial_2\cdot{\cal{K}}=\partial\cdot{\cal{K}}-H^2x{\cal{K}}'-\frac{1}{2}H^2D_1{\cal{K}}', \end{equation}
$D_1=H^{-2}\bar{\partial}$ and ${\cal{K}}'$ is the trace of ${\cal{K}}_{\alpha\beta}$. Now, utilizing the identities \cite{Garidi3838,Gazeau5920}
\begin{eqnarray}\label{identities}
\partial_2\cdot D_2{\Lambda_g}=-(Q_1+6){\Lambda_g},\;\; Q_2D_2{\Lambda_g}=D_2Q_1{\Lambda_g},
\end{eqnarray}
in which the action of the Casimir operator $Q_1$ on a vector field $\Lambda_g$ is
\begin{eqnarray}\label{identities'}
Q_1{\Lambda_g}=(Q_0-2){\Lambda_g}+2x \partial\cdot{\Lambda_g}-2\partial x\cdot{\Lambda_g},
\end{eqnarray}
the gauge invariance (\ref{2.4}) of the field equation can be easily presented as \cite{Gazeau+yo},
\begin{equation}\label{gauge} {\cal{K}}\rightarrow{\cal{K}}+D_2{\Lambda_g}, \end{equation}
while the gauge conditions (\ref{2.5}) takes the form
\begin{equation}\label{2.23}
\partial_2\cdot{\cal{K}}=({b}-\frac{1}{2})\bar{\partial}{\cal{K}}'.
\end{equation}

Respecting the field equation (\ref{2.21}), the corresponding action then would be
\begin{equation}\label{2.22}
S=\int d\sigma{\cal{L}},\;\;{\cal{L}}=-\frac{1}{2x^2}{\cal{K}}\cdot\cdot(Q_2+6){\cal{K}}+\frac{1}{2}(\partial_2\cdot{\cal{K}})^2.
\end{equation}
$d\sigma$ is the volume element in de Sitter space. By adding a gauge-fixing term to the Lagrangian, one obtains
\begin{eqnarray}\label{2.24}
{\cal{L}} & = & -\frac{1}{2x^2}{\cal{K}}\cdot\cdot(Q_2+6){\cal{K}}+\frac{1}{2}(\partial_2\cdot{\cal{K}})^2 \nn\\ &-&\frac{1}{2{a}}\Big( \partial_2\cdot{\cal{K}}-(b - \frac{1}{2})\bar{\partial}{\cal{K}}' \Big)^2.
\end{eqnarray}
Note that `$a$' and `$b$' are gauge-fixing parameters. Then, the variation of ${\cal{L}}$ leads to the equation
\begin{eqnarray}\label{2.26}
&(Q_2+6){\cal{K}}+ D_2\partial_2\cdot{\cal{K}}&\nn\\
&- \frac{1}{{a}}\Big( D_2\partial_2\cdot{\cal{K}} - (b-\frac{1}{2})^2 {\cal{S}} D_1\bar\partial{\cal{K}}'& \nn\\
& - {(b-\frac{1}{2})} (D_2\bar\partial{\cal{K}}' - {\cal{S}}D_1\partial_2\cdot{\cal{K}}) \Big)=0.&\hspace{1cm}
\end{eqnarray}

The general solution of the field equation (\ref{2.26}) can be constructed from a combination of a transverse-traceless (spin-two) part, ${\cal{K}}^{TT}$, plus a pure-trace (spin-zero) part, ${\cal{K}}^{PT}$,
$${\cal{K}}(x) = {\cal{K}}^{TT}(x) + {\cal{K}}^{PT}(x).$$
With respect to the Lagrangian (\ref{2.24}) and field equation (\ref{2.26}), therefore we have
\begin{eqnarray}\label{2.26TT}
(Q_2+6){\cal{K}}^{TT}+ (1 - \frac{1}{{a}}) D_2\partial_2\cdot{\cal{K}}^{TT} = 0,
\end{eqnarray}
and
\begin{eqnarray}\label{2.26S}
&(Q_2+6){\cal{K}}^{PT}+ (1 - \frac{1}{{a}}) D_2\partial_2\cdot{\cal{K}}^{PT} &\nn\\
& - \frac{1}{{a}}\Big(-(b-\frac{1}{2})^2 {\cal{S}} D_1\bar\partial{\cal{K}}' & \nn\\
& - {(b-\frac{1}{2})} (D_2\bar\partial{\cal{K}}' - {\cal{S}}D_1\partial_2\cdot{\cal{K}}^{PT} ) \Big)=0.&\hspace{1cm}
\end{eqnarray}
Finding the optimum value of `$a$' and `$b$' is practically a nontrivial question in curved spacetimes and have significant consequences on the two-point function that will be developed in the next sections.

\section{The transverse-traceless (spin-two) sector}
\subsection{Group theoretical content}
The elementary particle fields are classified by their corresponding UIR \`{a} \emph{la} Wigner.

Now, we explain that equation (\ref{2.26TT}) has a clear group theoretical content. The operator $Q_2$ commutes with the action of the de Sitter group generators and, therefore, it is constant in the corresponding UIR; the UIR's are classified by the use of eigenvalues of $Q_2$, i.e., $\langle Q_2\rangle$. According to Takahashi and Dixmier's notation \cite{Dixmier9,Takahashi289}, the eigenvalues of the Casimir operator,
$$\langle Q_p\rangle = -p(p+1) - (q+1)(q-2),$$
are classified under the following series representations in the present situation:
\begin{itemize}
\item{For Principal series representations $(U^{2,\nu})$ (also called ``massive" representations) \cite{Gazeau304008,Flato415}
\begin{equation}\label{2.30} \langle Q_2\rangle = \nu^2 - \frac{15}{4},\;\; p=2,\;q=\frac{1}{2}+i\nu;\; \nu\in\Re.  \end{equation}}
\item{For Complementary series representations $(V^{2,\mu})$
\begin{equation}\label{2.31} \langle Q_2\rangle = \mu - 4, \;\; p=2,\;q=\frac{1}{2}+\mu;\; \mu\in\Re,\;0<|\mu|<\frac{1}{2}.  \end{equation}}
\item{For Discrete series representations $({\Pi}^{\pm}_{2,q})$ (also called the ``massless" representations) \cite{Gazeau304008,Flato415},
\begin{equation}\label{2.32} \langle Q_2\rangle = - 6 - (q+1)(q-2), \;\;p=2,\; q={1},{2}.  \end{equation}
For the discrete series, regarding the parameter $q=1$ ($\langle Q_2\rangle =-4$), leads to the representation ${\Pi}^{\pm}_{2,1}$, which has no corresponding counterpart in the Minkowskian limit. The second value, $q=2$ ($\langle Q_2\rangle =-6$), however, leads to the representation ${\Pi}^{\pm}_{2,2}$. They are exactly the unique extensions of the massless Poincar\'{e} group representations with helicity $\pm2$.}
\end{itemize}

On this basis, the field equation for a transverse-traceless rank-$2$ tensor (or spin-$2$) field would be \cite{Garidi3838}
\begin{equation}\label{2.27}
(Q_2 - \langle Q_2\rangle){\cal{K}}^{TT}(x)=0.
\end{equation}
Constrained with the condition $\partial\cdot{\cal{K}}^{TT}=0$, this equation was solved in Ref. \cite{Garidi3838} rendering the following solution
\begin{eqnarray}\label{2222.26S}
{\cal{K}}^{TT} = \Big(-\frac{2}{3}\theta Z_1\cdot + {\cal{S}}\bar{Z}_1 + \frac{1}{\langle Q_2\rangle + 6} D_2[Z_1\cdot\bar\partial \nn\\
- H^2 xZ_1\cdot + 3H^2x\cdot Z_1 - \frac{1}{3}H^2D_1Z_1\cdot]\Big)K,
\end{eqnarray}
where $Z_1 (= Z_{1\alpha})$ is a five-dimensional constant vector ($\bar Z_{1\alpha}= \theta_{\alpha\beta}Z_1^\beta$) and $K$ is a vector field
\begin{equation}\label{222.27}
(Q_1 - \langle Q_1\rangle)K=0,
\end{equation}
with $\langle Q_1\rangle =  \langle Q_2\rangle + 4,\; x\cdot K=\partial\cdot K=0$.

Clearly, for the spin-$2$ massless field, Eq. (\ref{2222.26S}) reveals that the value $\langle Q_2\rangle= -6$ results in a singularity. This singularity is actually due to the divergencelessness condition needed to associate the tensor field with a specific UIR of the dS group. Therefore, the subspace specified by $\partial\cdot{\cal{K}}^{TT}=0$ considered so far is not sufficient for the construction of the massless tensor field. In order to suppress this difficulty, the divergencelessness condition must be dropped \cite{Garidi3838}. As a result two consequences follow immediately:
\begin{itemize}
\item{the appearance of gauge invariance in the field equation (see Eq. (\ref{2.26TT})), and}
\item{the necessity of using an indecomposable representation of dS group.\footnote{More precisely, in this context, massive elementary systems are associated with UIRs of the dS group \cite{Garidi3838}, while, massless elementary systems are connected to the indecomposable representations of this group \cite{Garidi032501,Gazeau329}.}}
\end{itemize}
The quantization of the tensor field, however, necessitates the fixing of the gauge parameter. This fixing bears the elimination of the singularity.
In the context of the de Sitter group theory, it is proved that the minimal (or optimal) choice, that restricts the space of solutions to the minimal content of any massless invariant theory, is \cite{Gazeau2533,Gazeau507}
\begin{equation}\label{c}c\;(=\frac{a-1}{a}) = \frac{2}{2s + 1},\end{equation}
$s$ is the angular momentum, spin, of the field. Any other choice of `$c$' represents logarithmic singularities, which implies reverberation inside the light cone \cite{Gazeau329}. Interestingly, investigating the massless vector field in dS space has proved that, in complete agreement with the general formula (\ref{c}), the minimal choice, for which no logarithmic-divergent terms appear, is $c=\frac{2}{3}$ \cite{Garidi032501}.

Pursuing this path, in the following section we present the solution of Eq. (\ref{2.26TT}) and in consistency with the above statements, we show that the optimal choice, for which the logarithmic contribution disappears, is $c=\frac{{a}-1}{{a}}=\frac{2}{5}$.

\subsection{Solution of the field equation}
We now solve the field equation (\ref{2.26TT}). Consider the traceless-transverse tensor field ${\cal{K}}^{TT}$, the most general solution, in terms of a five-dimensional constant vector $Z_1 (= Z_{1\alpha})$, a scalar field $\phi_1$ and two vector fields $K$ and $K_g$ by \cite{Gazeau2533}
\begin{equation} \label{4.1}
{\cal{K}}^{TT}=\theta\phi_1+ {\cal{S}}\bar Z_{1}K+D_{2}K_{g},
\end{equation}
where $x\cdot K= 0 = x\cdot K_g$ and
\begin{eqnarray}\label{4.2}
2\phi_1 + Z_{1}\cdot K + H^{-2}\bar\partial\cdot K_g = 0,
\end{eqnarray}
that is obtained from the tracelessness condition on (\ref{4.1}). We also impose the vector field $K$ to be divergenceless $\partial\cdot K= 0$.\footnote{Note that, for transverse tensors like $K$; $\partial\cdot K = \bar\partial\cdot K$.}

Applying (\ref{2.26TT}) to the above ansatz (\ref{4.1}) we obtain ($c= \frac{{a}-1}{{a}}$)
\begin{eqnarray}\label{4.3}
\left \{ \begin{array}{rl} (Q_0+6)\phi_1&=-4Z_1.K,\,\hspace{1.6cm}{(I)}\vspace{2mm}\\
\vspace{2mm} (Q_1+2)K &=Q_0K = 0, \hspace{0.8cm} {(II)}\\
\vspace{2mm}(Q_1+6)K_g&= \frac{c}{2(c-1)} H^2 D_1 \phi_1 \\ \vspace{2mm}&+ \frac{2-5c}{1-c}H^2x\cdot Z_1 K \\ \vspace{2mm}&+ \frac{c}{1-c}(H^2xZ_1\cdot K \\ \vspace{2mm}& \hspace{1.2cm}- Z_1\cdot\bar\partial K),\hspace{0.4cm}(III)\end{array}\right.
\end{eqnarray}
In order to obtain these equations we made use of the following commutation relations \cite{Gazeau507}
$$Q_2 D_2 K_g=D_2Q_1K_g,\;\;Q_2\theta\phi=\theta Q_0\phi,$$
$$Q_2S\bar Z_1 K = S\bar Z_1 (Q_1-4)K - 2H^2D_2x\cdot Z_1 K + 4\theta Z_1\cdot K,$$
$$\partial_2 \cdot \theta \phi = -H^2D_1\phi,\;\; \partial_2\cdot D_2K_g = -(Q_1+6)K_g.$$
Using the spectral theorem and the equations (\ref{4.3}-I) and (\ref{4.3}-II), the scalar field $\phi_1$ is completely determined by the vector field $K$ by the simple relation
\begin{eqnarray}\label{4.44444}
\phi_1 = -\frac{2}{3}Z_1\cdot K,
\end{eqnarray}
which also implies that $\phi_1$ verifies the massless minimally coupled scalar field equation $Q_0\phi_1 = 0$.

Now let us solve equation (\ref{4.3}-II). The most general form for the vector field $K$ is \cite{Gazeau2533}
\begin{equation}\label{4.5} K=\bar Z_2 \phi_2+D_1 \phi_3,\end{equation}
in which $\phi_2$ and $\phi_3$ are two scalar fields, and $Z_2$ is another 5-dimensional constant vector. By substituting (\ref{4.5}) into (\ref{4.3}-II) and using the condition $\bar\partial\cdot K = 0$, we have
\begin{equation} \label{4.6} \phi_3 =-\frac{1}{2}[Z_2.\bar\partial\phi_2 + 2H^2x.Z_2\phi_2], \end{equation}
\begin{equation} \label{4.7} Q_0 \phi_2=0. \end{equation}
This means that also $\phi_2$ verifies a massless minimally coupled scalar field equation. The vector field $K$, therefore, would be
\begin{equation} \label{4.8}
K=\bar Z_{2}\phi_2 - \frac{1}{2} D_1 [Z_2\cdot\bar\partial\phi_2 + 2H^2 x\cdot Z_2\phi_2],
\end{equation}
and from (\ref{4.44444}) we have
\begin{equation} \label{4.9}
\phi_1 =-\frac{2}{3}{Z_1}\cdot \Big( \bar Z_{2}\phi_2 - \frac{1}{2}D_1[Z_2\cdot\bar\partial\phi_2 + 2H^2x\cdot Z_2\phi_2]\Big).
\end{equation}

Now we will show that the vector field $K_g$ can also be obtained from the vector field $K$. In order to invert equation (\ref{4.3}-III) we will use the following identities \cite{Dehghani064028}
\begin{equation} \label{4.12} (Q_1+6) D_1 (Z_1\cdot K) = 6 D_1 (Z_1\cdot K),\end{equation}
\begin{equation} \label{4.13} (Q_1+6) x (Z_1\cdot K) = 6 x (Z_1\cdot K),\end{equation}
\begin{equation} \label{4.11} (Q_1+6)Z_1\cdot\bar\partial K = 6Z_1\cdot\bar\partial K+ 2H^2 D_1 (Z_1\cdot K),\end{equation}
\begin{equation} \label{4.10'} (Q_1+6)[H^2(x\cdot Z_1)K] = 2\Big[ H^2x(Z_1\cdot K) - Z_1\cdot\bar\partial K \Big],\end{equation}
$K_g$ is then obtained as
\begin{eqnarray} \label{4.14}
K_g  =  \frac{c}{2(1-c)}\Big[ H^2(x\cdot Z_1)K + \frac{1}{9}H^2D_1 (Z_1\cdot K) \Big] \nn\\
+ \frac{2-5c}{1-c} (Q_1 + 6)^{-1} H^2x\cdot Z_1 K  + \Lambda_g.
\end{eqnarray}
Note that $\Lambda_g$ is a vector field
$$(Q_1+6)\Lambda_g=0,\;\; x\cdot\Lambda_g=0,\;\; \bar\partial\cdot\Lambda_g=0.$$
Now our task is to handle $(Q_1 + 6)^{-1} H^2x\cdot Z_1 K$. In this regard, we utilize the plane wave formalism \cite{JBros1746,JBros327} and explicitly show that this term leads to a singularity in the solution.

Equation (\ref{4.7}) means that the scalar field $\phi_2$ obeys
\begin{equation} \label{4.777} \Box_H \phi_2 =0,\end{equation}
and its solutions are known to be the dS massless waves \cite{JBros1746,JBros327}
\begin{equation}\label{pwmmc} \phi_2 = (Hx\cdot\xi)^\sigma,\;\;\; \sigma=0,-3\end{equation}
where this 5-vector $\xi$ lies on the positive null cone ${\cal C}^{+} = \{ \xi \in \Re^5;\;\;\xi^2=0,\; {\xi}^0>0 \}$. Then, substituting (\ref{pwmmc}) into Eq. (\ref{4.8}) leads to
\begin{equation} \label{4.15} K = -\frac{\sigma}{2}\Big[\bar Z_{2} + (\sigma + 2)\frac{(x\cdot Z_2)}{(x\cdot\xi)}\bar\xi\Big]\phi_2. \end{equation}
Note that for simplicity the conditions $Z_1 \cdot \xi = Z_2 \cdot \xi = 0$ are imposed. Because of these conditions, the degree of freedom of 5-vectors $Z_1$ and $Z_2$ is reduced from 5 to 4. Using Eqs. (\ref{4.10'}) and (\ref{4.15}), we can easily show that
\begin{eqnarray} \label{4.16..}
(Q_1 + 6) H^2(x\cdot Z_1) K = -2\sigma H^2(x\cdot Z_1) K \hspace{2cm} \nn\\
+ \sigma H^2 \Big( (\sigma+3)(x\cdot Z_2)\bar Z_{1} + (\sigma + 2)H^{-2}\frac{Z_1\cdot Z_2}{x\cdot\xi}\bar\xi \Big)\phi_2,\hspace{0.5cm}
\end{eqnarray}
or equivalently
\begin{eqnarray} \label{equ}
(Q_1 + 6)^{-1} H^2(x\cdot Z_1) K = -\frac{1}{2\sigma} H^2(x\cdot Z_1) K \hspace{1cm} \nn\\
+ \frac{1}{2} (Q_1 + 6)^{-1}\Big[ H^2 \Big( (\sigma+3)(x\cdot Z_2)\bar Z_{1}\hspace{1cm} \nn\\
\hspace{2cm}+ (\sigma + 2)H^{-2}\frac{Z_1\cdot Z_2}{x\cdot\xi}\bar\xi \Big)\phi_2 \Big], \;
\end{eqnarray}
in which $\sigma = 0, -3$. Obviously, in order to handle $(Q_1 + 6)^{-1} H^2x\cdot Z_1 K$, we inevitably face an expression proportional to $\frac{1}{\sigma}$ which is divergent at $\sigma = 0$. We must, therefore, set $c = 2/5$ to eliminate the divergent solutions from (\ref{4.14}). On this basis, choosing $c=2/5$, $K_g$ would be
\begin{eqnarray} \label{4.18}
{K}_g^{(\frac{2}{5})}  =  \frac{1}{3}\Big[ H^2(x\cdot Z_1)K + \frac{1}{9}H^2D_1 (Z_1\cdot K) \Big].
\end{eqnarray}

Accordingly, using Eqs. (\ref{4.8}), (\ref{4.9}), and (\ref{4.18}), the field solution for $c=2/5$, ${{\cal{K}}}^{TT{(\frac{2}{5})}}$, can be written in the following form
\begin{equation} \label{4.19} {{\cal{K}}}_{\alpha \beta}^{TT{(\frac{2}{5})}} = {{\cal{D}}}^{TT{(\frac{2}{5})}}_{\alpha \beta}(x,\partial,Z_1,Z_2)\phi_2, \end{equation}
where $\phi_2\equiv\phi_m$ is the dS MMC scalar field and ${{\cal{D}}}^{TT{(\frac{2}{5})}}$ is the projection tensor
\begin{eqnarray}
{{\cal{D}}}^{TT{(\frac{2}{5})}}(x,\partial,Z_1,Z_2) = \Big( -\frac{2}{3} \theta Z_1\cdot +{\cal S}\bar Z_1 \hspace{2cm} \nn\\
+ \frac{1}{3}D_2 \Big[H^2(x\cdot Z_1) + \frac{1}{9}H^2D_1(Z_1\cdot )\Big]\Big) \nn\\
\times \Big( \bar Z_{2} - \frac{1}{2}D_1[(Z_2\cdot \bar\partial) + 2H^2(x\cdot Z_2)] \Big).\hspace{0.5cm}
\end{eqnarray}
We are now in the position to write the general field solution in the convenient following form
\begin{equation}
{\cal{K}}^{TT} = {{\cal{K}}}^{TT{(\frac{2}{5})}} + \frac{\frac{2}{5} - c}{1-c} D_2 (Q_1 + 6)^{-1} (\partial\cdot {{\cal{K}}}^{TT{(\frac{2}{5})}}).
\end{equation}
The term $(Q_1 + 6)^{-1} (\partial\cdot {{\cal{K}}}^{TT{(\frac{2}{5})}})$ is responsible for the singularity which implies reverberation inside the light cone (for more detailed discussions, see \cite{Gazeau329}). From now on, we shall work essentially with the $c=2/5$ gauge. It is actually the so-called \enquote{minimal case} in the context of de Sitter group theory \cite{Gazeau2533,Gazeau507}. Here, it is worth mentioning that the obtained result for the gauge-fixing parameter, $c=2/5$ and consequently ${a}=5/3$, is exactly what obtained in \cite{AHiguchi4317,AHiguchi124006} (it eliminates logarithmically divergent solutions).

Searching for a covariant quantization, one must find the minimum space of solutions that is dS-invariant. In this regard, we examine the dS covariance of the general solution by applying the action of the dS group on (\ref{4.19}). One can easily show that (Note that, for the sake of simplicity, from now on the index `$(\frac{2}{5})$' is omitted.)
\begin{eqnarray} \label{6996}
L_{\alpha\beta}{\cal{K}}_{\rho\delta}^{TT} = L_{\alpha\beta}\Big({\cal{D}}^{TT}_{\rho\delta}\phi_m\Big) = {\cal{D}}^{TT}_{\rho\delta}\Big(M_{\alpha\beta}\phi_m\Big).
\end{eqnarray}
Respecting explicit computation given in Ref. \cite{Gazeau1415}, considering any complete set of (positive norm) modes including the zero mode, one can easily see that the invariance of the transverse-traceless sector of the field solution (as for any gauge field) is broken owing to the structure function $\phi_m$. In fact, it is proved that \cite{Gazeau1415,de Bievre6230} the smallest complete, non-degenerate and invariant inner-product space for the MMC scalar field is a Krein space; a direct sum of a Hilbert space and an anti-Hilbert space (a space with definite negative inner product). To manage this difficulty and also the aforementioned gaugelike symmetry $\phi_m \longrightarrow \phi_m + \mbox{`constant'}$, as already pointed out, a canonical quantization method \`{a} \emph{la} Gupta-Bleuler in which the Fock space is built over the Krein space, the so-called Krein-Gupta-Bleuler quantization method, should be in order \cite{Gazeau1415,de Bievre6230,Garidi,Bertola}.

In the KGB context, the MMC scalar field operator $\varphi_m$ would be \cite{Gazeau1415,de Bievre6230,Garidi,Bertola}\footnote{Here, in order to simplify the notation, we consider ${\cal{P}}$ to be the set of indices for the positive norm modes
$$ {\cal{P}} = \{ (L,l,m) \in \mathds{N} \times \mathds{N} \times \mathds{Z};\; 0 \leq l \leq L, -l \leq m \leq l \}.$$}
\begin{eqnarray} \label{123}
\varphi_m = \frac{1}{\sqrt{2}} \Big( \sum_{{p} \in {\cal{P}}} [a_{p} (\phi_m)_{p} + a^\dagger_{p} (\phi^*_m)_{p}]\hspace{2cm}\nn\\
\hspace{2cm}+ \sum_{{p} \in {\cal{P}}} [b^\dagger_{p} (\phi_m)_{p} + b_{p} (\phi^*_m)_{p}] \Big).\;
\end{eqnarray}
Note that, the first sum on the right is the standard scalar field operator as was used by Allen \cite{BAllen3136,BAllen3771}. Then, we have the following operational relations
\begin{eqnarray}
& a_p |0\rangle=0,\;\;[a_p, a^\dagger_{p'}] = \delta_{p p'}, &\nn\\
& b_p |0\rangle=0,\;\;[b_p, b^\dagger_{p'}] = -\delta_{p p'}, &
\end{eqnarray}
other commutation relations are zero.

To review the Krein-Gupta-Bleuler lying behind the dS MMC and how the method uses a bigger Fock space on which negative norms are allowed and upon which acts a quantum field, more exactly, the properties of its vacuum, how observables are determined in this formalism, positivity of expectation values of the energy operator in all physical states which guarantees a reasonable physical interpretation of the theory (the so-called Wald axioms are well restored), and consistency of the method in the Minkowskian limit (unitarity conditon) one can refer to \cite{HPejhan044016}.\footnote{Also refer to \cite{Suren750,HPejhan1601} to see how the KGB method can be used for calculating physical observables e.g. the Casimir energy-momentum tensor in braneworld scenarios.}

Respecting the KGB quantization method, the corresponding transverse-traceless graviton two-point function is given in Sec. III-C. We show that the method enables us to obtain the fully dS-covariant and infrared-free construction for the linearized gravity.

\subsubsection{Comment on the gauge-fixing procedure in \cite{HPejhan044016}}
As already discussed, a group theoretical approach to massless fields in dS spacetime reveals that, in order to avoid logarithmic-divergent terms, the divergencelessness condition which is needed to associate such tensor fields with specific UIRs of the dS group must be dropped. The quantization procedure, therefore, necessitates the fixing of the gauge parameter. It is, however, reported in \cite{HPejhan044016} that exerting the extra condition $\bar{\partial} \cdot K = 0$, which contracts the solutions space, could suppress the logarithmic divergences from the theory. The conclusions of \cite{HPejhan044016} were based on calculations and reasonings presented in \cite{Dehghani064028}, where, in spite of imposing the divergencelessness condition on the dS massless spin-$2$ field, the corresponding two-point function is free of any logarithmic divergences. These calculations convinced us that applying the extra condition $\bar{\partial} \cdot K = 0$ contracts the space of solutions so that the logarithmic contribution disappears; it could have been supposed as the last option, by relaxing which, we could recover the existence of the logarithmic divergences anticipated by the group theory. Then, in order to prove it and extend our previous work \cite{HPejhan044016}, we decided to solve the equation without applying the extra condition to obtain the most general form for the two-point function and show that how the logarithmic divergences, in consistency with the content of the de Sitter group theory, appear. Therefore, we relaxed the $\bar{\partial} \cdot K = 0$ condition and performed calculations from the very beginning. Surprisingly we found, through tedious but straightforward calculations, that even by relaxing this condition no logarithmic divergence appears. Respecting to the de Sitter group theory (see part A. Group theoretical content) then rang a bell. Something was not right!

Investigating the whole procedure revealed that the shortcoming was concealed in the Appendix E expressions of the paper \cite{Dehghani064028}, more exactly how (E6) and consequently (E7) are obtained. Technically, the expressions (E1) to (E4) (the same as  the above (\ref{4.12}) to (\ref{4.10'})) are correct, and based on them one can easily write (E5)
\begin{eqnarray} \label{E5}
(Q_1 + 6)[(x \cdot Z_1)K] = \hspace{4 cm} \nn\\ \frac{1}{3}(Q_1 + 6)\Big[ \frac{1}{3}D_1(Z_1 \cdot K) + x(Z_1 \cdot K) - Z_1 \cdot \bar{\partial}K \Big].
\end{eqnarray}
To obtain (E6), the differential operator $(Q_1 + 6)$ is dropped from both sides as follows
\begin{eqnarray} \label{E6}
(x \cdot Z_1)K = \frac{1}{3}\Big[ \frac{1}{3}D_1(Z_1 \cdot K) + x(Z_1 \cdot K) - Z_1 \cdot \bar{\partial}K \Big], \hspace{2mm}
\end{eqnarray}
which is obviously an illegitimate action and indeed the origin of the problem. More precisely, this illegal procedure has automatically but incorrectly avoided the logarithmic divergences to appear in the theory.

On this basis, in the present paper, we correct the calculations given in our previous work \cite{HPejhan044016}; the correct form of Eq. (48) for $K_g$ in \cite{HPejhan044016} is as Eq. (\ref{4.18}). This also imposes another corrections to \cite{HPejhan044016} which will be noticed in appropriate places. However, it must be emphasized here that the aforementioned corrections do not alter the main achievements of \cite{HPejhan044016} (we will prove it in the rest of the present paper). More accurately, the final transverse-traceless graviton two-point function is free of IR divergences if and only if we use the Krein-Gupta-Bleuler (an indefinite inner product) quantization scheme. Otherwise, pursuing the standard quantization approach, naturally results in that dS-breaking (the appearance of IR divergences) is universal. Furthermore, by choosing two gauge-fixing parameters, the two-point function for the pure-trace part can still be written in terms of the massless conformally coupled scalar field (see Sec. IV).

\subsection{The two-point function}
Pursuing Allen and Jacobson procedure in reference \cite{allen2}, the two-point functions in de Sitter space are written in terms of bi-tensors in this section (bi-tensors are functions of two points $(x, x')$ and behave like tensors under coordinate transformations at each point \cite{allen2}). Bi-tensor two-point functions are the cornerstone of the dS axiomatic field theory construction \cite{JBros327}. Bi-tensors are called maximally symmetric if they hold dS invariance \cite{allen2}. Any maximally symmetric bi-tensor can be expressed in ambient space notations as
\begin{eqnarray} \label{WTT}
{\cal{W}}^{TT}_{\alpha\beta\alpha'\beta'}(x,x')= \theta \theta' {\cal{W}}_0(x,x') + {\cal{S}}{\cal{S}}'\theta\cdot\theta'{\cal{W}}_1(x,x')\nn\\
+D_2D'_2{\cal{W}}_g(x,x'),\;\;\;\;\;\;\;\;\;\;\;\;\;\;\;\;\;\;\;\;\;\end{eqnarray}
where ${\cal{W}}_1$ and ${\cal{W}}_g$ are transverse bi-vectors, ${\cal{W}}_0$ is bi-scalar and $D_2D'_2= D'_2D_2$. The two-point function must verify the field equation (\ref{2.26TT}), regarding both $x$ and $x'$ (with no difference). We first choose $x$ to start our study. By making bi-tensor (\ref{WTT}) to comply Eq. (\ref{2.26TT}), one finds ($c=\frac{2}{5}$)
\begin{widetext}
\begin{equation}\label{6.11}
\left\{\begin{array}{rl} &(Q_0+6)\theta'{\cal{W}}_0=-4{\cal S}'\theta'\cdot {\cal{W}}_{1},\hspace{1.7cm}(I)\vspace{2mm}\\
&(Q_1+2){\cal{W}}_{1}=0, \hspace{3.5cm}(II)\vspace{2mm}\\
&(Q_1+6)D'_2{\cal{W}}_g= \frac{-1}{3} H^2 D_1 \theta'{\cal{W}}_0+ H^2{\cal S}'\Big[\frac{2}{3}(D_1\theta'\cdot - x\theta'\cdot - H^{-2}\theta'\cdot\bar\partial)\Big]{\cal{W}}_{1}.\hspace{0.5cm}(III)
\end{array}\right.
\end{equation}
\end{widetext}
where the condition $\partial\cdot{\cal{W}}_1 = 0$, is exerted. Considering Eqs. (\ref{6.11}-I) and (\ref{6.11}-II) yields
\begin{equation} \label{6.12} \theta'{\cal{W}}_0(x,x')=-\frac{2}{3}{\cal S}'\theta'\cdot{\cal{W}}_{1}(x,x').\end{equation}
The bi-vector two-point function ${\cal{W}}_{1}$, which is the solution of Eq. (\ref{6.11}-II), can be written as
$${\cal{W}}_{1}=\theta\cdot\theta'{\cal{W}}_{2}+D_1D'_1{\cal{W}}_{3}.$$
where ${\cal{W}}_{2}$ and ${\cal{W}}_{3}$ are bi-scalar two-point functions, so that
$$D'_1{\cal{W}}_{3}=-\frac{1}{2}[2H^2 (x\cdot\theta'){\cal{W}}_2 + \theta'\cdot\bar\partial{\cal{W}}_{2}],$$
$$Q_0{\cal{W}}_{2}=0.$$
Therefore, ${\cal{W}}_{2}\equiv{\cal{W}}_{mc}$ is a MMC bi-scalar two-point function. Regarding the above identities, we will have the bi-vector two-point function as follows
\begin{equation}\label{6.13} {\cal{W}}_{1}(x,x')=\Big(\theta\cdot\theta' - \frac{1}{2}D_{1}[\theta'\cdot\bar\partial + 2H^2 x\cdot\theta']\Big){\cal{W}}_{mc}(x,x').\end{equation}

Following a similar procedure used to calculate (\ref{4.12}) to (\ref{4.10'}), we will obtain
$$ (Q_1+6) x\theta' \cdot {\cal{W}}_{1} = 6 x\theta' \cdot {\cal{W}}_{1},$$
$$ (Q_1+6)D_1 \theta'\cdot{\cal{W}}_{1} = 6 D_1 \theta'\cdot{\cal{W}}_{1}, $$
$$ (Q_1+6)\theta'\cdot\bar\partial{\cal{W}}_{1} = 6 \theta'\cdot\bar\partial{\cal{W}}_{1} + 2H^2 D_1 (\theta'\cdot{\cal{W}}_{1}), $$
$$ (Q_1+6)[H^2(x\cdot\theta'){\cal{W}}_{1}] = 2\Big[H^2 x(\theta'\cdot{\cal{W}}_{1}) - (\theta'\cdot\bar\partial){\cal{W}}_{1} \Big].$$
Utilizing the above identities together with Eqs. (\ref{6.11}-III) and (\ref{6.12}), we have
\begin{eqnarray}\label{6.14}
D'_2 {\cal{W}}_{g}(x,x') =\frac{H^2}{3}{\cal S}' \Big[x\cdot\theta'{\cal{W}}_{1} + \frac{1}{9}D_1 \theta'\cdot{\cal{W}}_{1} \Big].
\end{eqnarray}

Correspondingly, the bi-tensor two-point function (\ref{WTT}) will be
\begin{equation} \label{6.18'} {\cal W}^{TT}_{\alpha\beta\alpha'\beta'}(x,x')= \Delta^{TT}_{\alpha\beta\alpha'\beta'}(x,x'){\cal W}_{mc}(x,x'), \end{equation}
where
\begin{widetext}
\begin{eqnarray}\label{6.19'}
\Delta^{TT} (x,x')= -\frac{2}{3}{\cal S}'\theta\theta'\cdot \Big(\theta\cdot\theta' - \frac{1}{2}D_{1}[\theta'\cdot\bar\partial + 2H^2 x\cdot\theta']\Big) +{\cal S}{\cal S}'\theta\cdot\theta'\Big(\theta\cdot\theta' - \frac{1}{2}D_{1}[\theta'\cdot\bar\partial + 2H^2 x\cdot\theta']\Big)\nn\\
+\frac{1}{3}H^2{\cal S}'D_2 \Big[x\cdot\theta' + \frac{1}{9}D_1 \theta'\cdot \Big]\Big(\theta\cdot\theta' - \frac{1}{2}D_{1}[\theta'\cdot\bar\partial + 2H^2 x\cdot\theta']\Big).
\end{eqnarray}

Furthermore, the bi-tensor (\ref{WTT}) must verify Eq. (\ref{2.26TT}) with respect to $x'$. So, following the same routine, we obtain
$$ \left\{\begin{array}{rl} &(Q'_0+6)\theta{\cal{W}}_0=-4{\cal S}\theta\cdot{\cal{W}}_{1},\hspace{1.9cm}(I)\vspace{2mm}\\
&(Q'_1+2){\cal{W}}_{1}=0, \hspace{3.5cm}(II)\vspace{2mm}\\
&(Q'_1+6)D_2{\cal{W}}_g= -\frac{1}{3} H^2 D'_1 \theta{\cal{W}}_0 + H^2{\cal S}\Big[\frac{2}{3}(D'_1\theta\cdot - x'\theta\cdot - H^{-2}\theta\cdot\bar\partial')\Big]{\cal{W}}_{1}.\hspace{1cm}(III)
\end{array}\right.$$
here, the condition $\partial'\cdot{\cal{W}}_1 = 0$ is applied. In this situation, we have
\begin{equation} \label{6.15} \theta{\cal{W}}_0(x,x')= -\frac{2}{3}{\cal S}\theta\cdot{\cal{W}}_{1}(x,x'),\end{equation}
\begin{equation}\label{6.16} {\cal{W}}_{1}(x,x')=\Big(\theta\cdot\theta' - \frac{1}{2}D'_{1}[\theta\cdot\bar\partial' + 2H^2 x'\cdot\theta]\Big){\cal{W}}_{mc}(x,x'),\end{equation}
\begin{eqnarray}\label{6.17}
D_2 {\cal{W}}_{g}(x,x') = \frac{H^2}{3}{\cal S}\Big[ x'\cdot\theta{\cal{W}}_{1} + \frac{1}{9}D'_1 \theta\cdot{\cal{W}}_{1}\Big].
\end{eqnarray}

Utilizing Eqs. (\ref{6.15})-(\ref{6.17}), we can write the bi-tensor two-point function in the following form
\begin{equation} \label{6.18} {\cal{W}}^{TT}_{\alpha\beta\alpha'\beta'}(x,x')= \Delta'^{TT}_{\alpha\beta\alpha'\beta'}(x,x'){\cal{W}}_{mc}(x,x'), \end{equation}
where
\begin{eqnarray}\label{6.19}
\Delta'^{TT} (x,x')= -\frac{2}{3}{\cal S}\theta'\theta\cdot \Big(\theta'\cdot\theta - \frac{1}{2}D'_{1}[\theta\cdot\bar\partial' + 2H^2 x'\cdot\theta]\Big) +{\cal S}{\cal S}'\theta'\cdot\theta\Big(\theta'\cdot\theta - \frac{1}{2}D'_{1}[\theta\cdot\bar\partial' + 2H^2 x'\cdot\theta]\Big)\nn\\
+\frac{1}{3}H^2{\cal S}D'_{2} \Big[x'\cdot\theta + \frac{1}{9}D'_1 \theta\cdot \Big]\Big(\theta'\cdot\theta - \frac{1}{2}D'_{1}[\theta\cdot\bar\partial' + 2H^2 x'\cdot\theta]\Big).
\end{eqnarray}

Briefly, till here by utilizing an ansatz similar to the one used for computing the field solutions, we have shown that transverse-traceless sector of the graviton two-point function can be written in the sense of the MMC scalar two-point function ${\cal{W}}_{mc}(x,x')$. This must be underlined here that the fundamental assumption that the whole procedure fulfills de Sitter invariance is the base of our calculations from the very beginning (see (\ref{WTT})). More exactly, only upon this assumption, the transverse-traceless graviton two-point function can be written in terms of the maximally symmetric bi-tensors \cite{allen2}. As already pointed out, the only way to preserve dS invariance is including all the negative frequency solutions in the theory. Indeed, the Krein-Gupta-Bleuler construction should be in order \cite{Gazeau1415,de Bievre6230,Garidi,Bertola}. This results in that ${\cal{W}}_{mc}$ is only a function of the invariant length ${\cal{Z}}\equiv -H^2x\cdot x'$; ${\cal W}_{mc} = {\cal W}_{mc}({\cal Z})$, and hence, the equation $Q_0{\cal{W}}_{mc}({\cal{Z}})=0$ turns into the ordinary differential equation
\begin{equation} \label{WCMMC}
\Big( (1-{\cal{Z}}^2)\frac{d^2}{d{\cal{Z}}^2} - 4{\cal{Z}} \frac{d}{d{\cal{Z}}} \Big){\cal{W}}_{mc}({\cal{Z}}) = 0.
\end{equation}

Now, we can use the formulas given in \cite{HPejhan044016} (see App. A) to obtain the following expressions
\begin{equation} \label{W0MMC} \theta'_{\alpha'\beta'}{\cal W}_{0}(x,x')=  \frac{1}{3}{\cal S}' \Big[\theta'_{\alpha'\beta'} + \frac{4}{1-{\cal{Z}}^2}H^2 (x\cdot\theta'_{\alpha'})(x\cdot\theta'_{\beta'})\Big]{\cal{Z}}\frac{d}{d{\cal{Z}}}{\cal W}_{mc}({\cal{Z}}),
\end{equation}
\begin{equation} \label{W1MMC}
{\cal W}_{1\beta\beta'}(x,x')=  \frac{1}{2} \Big[\frac{3+{\cal{Z}}^2}{1-{\cal{Z}}^2}H^2 (x'\cdot\theta_{\beta})(x\cdot\theta'_{\beta'}) - {\cal{Z}} (\theta_{\beta}\cdot\theta'_{\beta'})\Big]\frac{d}{d{\cal{Z}}}{\cal W}_{mc}({\cal{Z}}),
\end{equation}
\begin{eqnarray}\label{WgMMC}
D_{2\alpha}D'_{2\alpha'}{\cal W}_{g\beta\beta'}(x,x') = & - & \frac{H^2}{54(1-{\cal{Z}}^2)^2}{\cal S}{\cal S}'\Big[ H^{-2}{\cal{Z}}(1-{\cal{Z}}^2)( 1+3{\cal{Z}}^2 ) \theta_{\alpha\beta} \theta'_{\alpha'\beta'} \nn\\ & + & H^{-2}{\cal{Z}}(1-{\cal{Z}}^2)( 17-9{\cal{Z}}^2 )(\theta_{\alpha}\cdot\theta'_{\alpha'}) (\theta_{\beta}\cdot\theta'_{\beta'}) + 24{\cal{Z}}( 2-{\cal{Z}}^2 )\theta_{\alpha\beta}(x\cdot\theta'_{\alpha'})(x\cdot\theta'_{\beta'}) \nn\\ & + &  ( -79-62{\cal{Z}}^2+45{\cal{Z}}^4 )(\theta_{\alpha}\cdot\theta'_{\alpha'}) (x\cdot\theta'_{\beta'}) (x'\cdot\theta_{\beta}) + 12{\cal{Z}}( 1+{\cal{Z}}^2 )\theta'_{\alpha'\beta'}(x'\cdot\theta_{\alpha})(x'\cdot\theta_{\beta}) \nn\\ & + & \frac{12{\cal{Z}}H^2}{1-{\cal{Z}}^2} ( 21-2{\cal{Z}}^2-3{\cal{Z}}^4 ) (x'\cdot\theta_{\alpha})(x'\cdot\theta_{\beta})(x\cdot\theta'_{\alpha'})(x\cdot\theta'_{\beta'}) \Big] \frac{d}{d{\cal{Z}}}{\cal W}_{mc}({\cal{Z}}).
\end{eqnarray}
By substituting Eqs. (\ref{W0MMC})-(\ref{WgMMC}) into (\ref{WTT}) we obtain the exact form of the two-point function (the Krein two-point function, which is actually the commutator \cite{HPejhan044016,Gazeau1415}) in the ambient space formalism as follows
\begin{eqnarray}\label{WMMC}
{\cal W}^{TT}_{\alpha\beta \alpha'\beta'}(x,x') & = & {\frac{2{\cal{Z}}}{27(1-{\cal{Z}}^2)^2}}{\cal S}{\cal S}' \nn\\
& \times & \Big[ \theta_{\alpha\beta}\theta'_{\alpha'\beta'}f_1({\cal{Z}}) + (\theta_{\alpha}\cdot\theta'_{\alpha'})(\theta_{\beta}\cdot\theta'_{\beta'})f_2({\cal{Z}})
+ H^2\Big(\theta'_{\alpha'\beta'}(x' \cdot \theta_{\alpha}) (x'\cdot\theta_{\beta}) + \theta_{\alpha\beta}(x\cdot\theta'_{\alpha'})(x\cdot\theta'_{\beta'}) \Big)f_3({\cal{Z}})\nn\\
& + & H^4\Big( (x'\cdot\theta_{\alpha})(x'\cdot\theta_{\beta})(x\cdot\theta'_{\alpha'})(x\cdot\theta'_{\beta'}) \Big)f_4({\cal{Z}})
+ (\theta_{\alpha}\cdot\theta'_{\alpha'})(x\cdot\theta'_{\beta'})(x'\cdot\theta_{\beta})f_5({\cal{Z}}) \Big] \frac{d}{d{\cal{Z}}}{\cal W}({\cal{Z}}),
\end{eqnarray}
in which
$$f_1({\cal Z})=(1-{\cal Z}^2) (2-3{{\cal{Z}}}^2),\;\;f_2({\cal{Z}}) = (1-{\cal{Z}}^2) (-11 + 9{{\cal{Z}}}^2),\;\;f_3({\cal{Z}}) = -3(1 + {{\cal{Z}}}^2),$$
$$f_4({\cal{Z}}) = - \frac{3}{(1-{\cal Z}^2)}( 21 - 2{{\cal{Z}}}^2 - 3{{\cal{Z}}}^4),\;\;f_5({\cal{Z}})= \frac{1}{{\cal{Z}}} ( 40+ 2{\cal{Z}}^2 - 18{\cal{Z}}^4 ).$$
\end{widetext}
\emph{Note:} In the above computations, it is presumed that the two points, $x$ and $x'$, are not located on the light cone of each other and hence $1- {\cal{Z}} \neq 0$.

Regarding the differential equation (\ref{WCMMC}), the function ${\cal{W}}_{mc}$ (the general solution) will be \cite{HPejhan044016}
\begin{equation} \label{WCMMC2}
{\cal{W}}_{mc}({\cal{Z}}) = C_1 \Big( \frac{1}{1 + {\cal{Z}}} - \frac{1}{1 - {\cal{Z}}} + \mbox{ln}\frac{1 - {\cal{Z}}}{1 + {\cal{Z}}} \Big) + C_2,
\end{equation}
where $C_1$ and $C_2$ are real constants. This function brings forward problems with locality \cite{Bertola}. However, this is not concerning because in the two-point function (\ref{WMMC}), this function enters only through its derivative,
\begin{equation} \label{WCMMC2'}
\frac{d}{d{\cal{Z}}}{\cal{W}}_{mc}({\cal{Z}}) = \frac{-4C_1}{({\cal{Z}}^2 - 1)^2},
\end{equation}
which is a local function. Now, by substituting (\ref{WCMMC2'}) into (\ref{WMMC}), because of the large order of ${\cal{Z}}$ in the denominator of (\ref{WCMMC2'}), it is easily seen that the large-distance growth of the two-point function (including linearly growing terms which appear due to the presence of expressions $f_1({\cal{Z}})$ and $f_2({\cal{Z}})$) clearly will not be reflected in the computed two-point function (\ref{WMMC}).

In summery, the two-point function (\ref{WMMC}) satisfies the following conditions:
\begin{itemize}
\item{\textbf{Indefinite sesquilinear form}\\
For any test function $f_{\alpha\beta}\in{\cal{D}}(X_H)$, an indefinite sesquilinear form is given by
\begin{equation}\label{6.4} \int_{X_H\times X_H}f^{\ast\alpha\beta}(x){\cal{W}}^{TT}_{\alpha\beta\alpha'\beta'}(x,x')f^{\alpha'\beta'}(x')d\sigma (x)d\sigma (x'), \end{equation}
where $f^\ast$ is the complex conjugate of $f$ and $d\sigma(x)$ characterizes the dS-invariant measure on $X_H$. ${\cal{D}}(X_H)$ is the space of functions $C^{\infty}$ with compact support in $X_H$.}

\item{\textbf{Locality}\\
For every space-like separated pair $(x,x')$, \textit{i.e.} $x\cdot x'>-H^{-2}$,
\begin{equation}\label{6.5} {\cal{W}}^{TT}_{\alpha\beta\alpha'\beta'}(x,x')={\cal{W}}^{TT}_{\alpha'\beta'\alpha\beta}(x',x). \end{equation}}

\item{\textbf{Covariance}
\begin{equation}\label{6.6} (g^{-1})_\alpha^\gamma(g^{-1})_\beta^\delta{\cal{W}}^{TT}_{\gamma \delta \gamma' \delta'}(gx,gx')g_{\alpha'}^{\gamma'}g_{\beta'}^{\delta'}={\cal{W}}^{TT}_{\alpha\beta\alpha'\beta'}(x,x'), \end{equation}
for all $g\in SO_0(1,4)$.}

\item{\textbf{Index symmetrizer}
\begin{equation}\label{6.7} {\cal{W}}^{TT}_{\alpha\beta\alpha'\beta'}(x,x')={\cal{W}}^{TT}_{\beta\alpha\beta'\alpha'}(x,x'). \end{equation}}

\item{\textbf{Transversality}
\begin{equation}\label{6.8} x^\alpha{\cal{W}}^{TT}_{\alpha\beta\alpha'\beta'}(x,x')=0=x'^{\alpha'}{\cal{W}}^{TT}_{\alpha\beta\alpha'\beta'}(x,x'). \end{equation}}

\item{\textbf{Tracelessness}
\begin{equation}\label{6.9} ({\cal{W}}^{TT})_{\;\;\;\alpha\alpha'\beta'}^\alpha(x,x')=0={({\cal{W}}^{TT})_{\alpha\beta\alpha'}}^{\alpha'}(x,x'). \end{equation}}
\end{itemize}

At the end of this part, we mention that the final spin-two part graviton two-point function in \cite{HPejhan044016} (Eq. (81) and underlying expressions) should be replaced by Eq. (\ref{WMMC}) without altering any conceptual discussions about the two-point function and its properties.

We complete this section by demonstrating the two-point function projected onto the de Sitter intrinsic space as follows (to review the details of projecting procedure one can refer to \cite{HPejhan044016})
\begin{eqnarray}\label{WMMCI} Q^{TT}_{\mu\nu\mu'\nu'}(X,X') & = & \frac{2{\cal{Z}}}{27}{\cal{S}}{\cal{S}}'\Big[  \frac{f_1}{(1-{\cal{Z}}^2)^2}g_{\mu\nu}g'_{\mu'\nu'} \nn\\ & + & \frac{f_2}{(1-{\cal{Z}}^2)^2}g_{\mu\mu'}g'_{\nu\nu'}\nn\\ & + & \frac{f_3}{1-{\cal{Z}}^2}(g_{\mu\nu}n_{\mu'}n_{\nu'}+g'_{\mu'\nu'}n_\mu n_\nu)\nn\\
& + & \Big( \frac{2({\cal{Z}}-1)f_2}{(1-{\cal{Z}}^2)^2}+\frac{f_5}{1-{\cal{Z}}^2} \Big)g_{\mu\mu'}n_\nu n_{\nu'}\nn\\
& + & \Big( \frac{f_2}{(1+{\cal{Z}})^2}-\frac{f_5}{1+{\cal{Z}}}+f_4 \Big)n_\mu n_\nu n_{\mu'}n_{\nu'} \Big]\nn\\
& \times & \frac{d}{d{\cal{Z}}}{\cal{W}}_{mc}({\cal{Z}}).\hspace{0.5cm}
\end{eqnarray}
The coefficients in this expansion are functions of the geodesic distance $\sigma(x, x')$ and the parallel propagator $g_{\mu\nu'}$,
$$ n_\mu = \nabla_\mu \sigma(x, x')\;\;\;,\;\;\; n_{\mu'} = \nabla_{\mu'} \sigma(x,x'),$$
$$ g_{\mu\nu'}=-c^{-1}({\cal{Z}})\nabla_{\mu}n_{\nu'}+n_\mu n_{\nu'}.$$
For $ {\cal{Z}}=-H^2x\cdot x'$, the geodesic distance can be characterized by
\begin{eqnarray}
\left\{\begin{array}{rl}
     {\cal{Z}}&=\cosh (H\sigma ),\hbox{if $x$ and $x'$ are time-like separated,} \\
     {\cal{Z}}&=\cos (H\sigma ),\hbox{if $x$ and $x'$are space-like separated.} \\
\end{array}\right.\;\;
\end{eqnarray}
The two-point function (\ref{WMMCI}) has been written completely in terms of ${\cal{Z}}$ in the dS global coordinate and hence is dS-invariant.\footnote{Note that, ${\cal{Z}}(x,x')$ is an invariant entity under the isometry group $O(1,4)$ and therefore any function constructed by ${\cal{Z}}$ is also dS-invariant.} It is also free of any infrared divergences. In order to see a comparison between our results and existing results in the literature, one can refer to App. A.

\emph{Concluding remarks:} We have shown that there exists no nontrivial covariant two-point function of the positive type for the spin-two sector of the gravitons field; this supports the statements by Woodard \emph{et al.} that ``\emph{including negative norm states in the mode sum is the only way to avoid de Sitter breaking}" (see Sec. I). More exactly, the only two-point function naturally appears is the Krein two-point function which is actually the commutator, but it is not of positive type and it does not allow to select physical states. Furthermore, our vacuum (the KGB vacuum) does not fit in the usual classification of vacua which is based on two-point functions. In this regard, we must insist on the fact that it is the field itself which is different in our construction and not only the vacuum \cite{HPejhan044016,de Bievre6230,Gazeau1415,Garidi}.

\section{The pure-trace (spin-zero) sector}
Thus far, the spin-two sector of linearized gravitons has been studied. It has been shown that through a standard linear covariant gauge-fixing procedure one can eliminate the logarithmic divergences associated with this part. Moreover, it has been proved that the transverse-traceless graviton two-point function is universally IR-divergent unless the KGB approach is considered.

In this section, we study the pure-trace (spin-zero) part, ${\cal{K}}^{PT}$, of the graviton two-point function. It should be noted that the spin-zero sector does not correspond to a UIR of the dS group. In fact, the tracelessness condition on the tensor field is an essential condition to relate it to the UIRs of the dS group \cite{Gazeau5920}.

To start, we consider
\begin{equation} \label{szs}
{\cal{K}}^{PT} = \frac{1}{4}\theta\Psi,
\end{equation}
where $\Psi$ is a scalar field. By taking trace of Eq. (\ref{2.26S}) and putting $\alpha = 5/3$, we can obtain the field equation for the scalar $\Psi$, which in order to be comparable with the results of \cite{HPejhan044016}, we demonstrate it as follows
\begin{equation} \label{PT}
\Big( Q_0 + \frac{12}{2 + 2f(b)} \Big)\Psi = 0,
\end{equation}
where $f(b)$ is a real number
\begin{equation} \label{fb}
f(b) = \frac{1}{5} - \frac{3}{2}(b-\frac{1}{2}) - \frac{6}{5}(b-\frac{1}{2})^2.
\end{equation}
On the other hand, any scalar field in accordance with the scalar discrete series UIR of the dS group obeys the subsequent equation with integer $n$ \cite{Dixmier9}
\begin{equation} \label{dsUIR}
(Q_0 + n(n+3))\Psi = 0.
\end{equation}

Well-known difficulties arise when we try to quantize these fields with the so-called ``imaginary mass" (with $2f(b)>-2$ or discrete series with $n>0$).
The corresponding two-point functions of these field exhibits a pathological large distance behavior \cite{BRatra1931}
\begin{equation} \label{TPFS}
{\cal{W}} \approx |{\cal{Z}}(x,x')|^{-\frac{3}{2} + \frac{\sqrt{9 + \frac{18}{2 + 2f(b)}}}{2}},
\end{equation}
for which, the choice $2f(b)<-2$ eliminates the pathological large distance behavior from the spin-zero sector two-point function. By applying this condition on Eq. (\ref{fb}), we find a rang for the gauge-fixing parameter `$b$' as follows
$$b>\frac{-3 + \sqrt{801}}{24}\;\;,\;\;b<\frac{-3 - \sqrt{801}}{24},$$
so that, de Sitter invariance is indeed restored and the theory is infrared-free. Therefore, the obtained result for the spin-zero sector is perfectly compatible with the usual point of view which supports that this sector is gauge-dependent and hence the introduced divergences can be eliminated by suitable gauge-fixing procedure \cite{PMora122502,AHiguchi4317,AHiguchi124006}.

On the other hand, by comparing Eq. (\ref{PT}) with (\ref{dsUIR}) while $2f(b)<-2$, one can easily show that only the values $n = -1, -2$, which render $2f(b) = -8$, relate Eq. (\ref{PT}) to the scalar series UIR of the dS group (see \cite{HPejhan044016}). By applying this condition on Eq. (\ref{fb}), one can obtain the ``optimal" value for `$b$'
$$b = \frac{-3 \pm \sqrt{2241}}{24},$$
which converts Eq. (\ref{PT}) to
\begin{equation} \label{dSMCC}
(Q_0 - 2)\Psi = 0.
\end{equation}
It, $\Psi$, is indeed the conformally coupled massless scaler field in de Sitter space.

\section{conclusion}
In this paper, through a rigorous group theoretical approach, we have obtained the fully dS-covariant graviton two-point function in dS space in a gauge with two parameters `$a$' and `$b$'. An appropriate gauge-fixing procedure enables us to eliminate logarithmic divergence and the dS-breaking contribution to the spin-zero part. Furthermore, in complete agreement with Woodard viewpoint (e.g. see \cite{RWoodard1430020} and references therein), we have proved that the de Sitter breaking of the transverse-traceless part of the linearized gravitons two-point function is gauge-independent and quite universal. We have also shown that the only way to eliminate IR divergences to preserve covariance of the theory is utilizing the Krein-Gupta-Bleuler quantization scheme which includes all the unphysical negative norm states in the theory and inevitably breaks the analyticity of the two-point function (the obtained Krein two-point function, actually, is the commutator).

Frankly speaking, in quantizing procedure of the linearized gravitons in de Sitter spacetime, covariance and analyticity cannot be summoned under one single roof.

\section*{Acknowledgements}
We would like to thank Professor J. P. Gazeau for remarkable comments and providing us with his unpublished notes. We also thank the referee for instrumental comments.

\begin{appendix}
\setcounter{equation}{0}
\begin{widetext}
\section{A comparison with existing results}
In this section, we compare the Krein two-point function (the commutator) (\ref{WMMCI}) with existing results. In this regard, first, by substituting the expressions for $d/d{\cal{Z}}{\cal{W}}_{mc}({\cal{Z}})$, Eq. (\ref{WCMMC2'}) (for the sake of simplicity we choose $C_1 = 1$), and $f_1$ to $f_5$ (below Eq. (\ref{WMMC})), the explicit form of the two-point function (\ref{WMMCI}) is obtained as follows
\begin{eqnarray} \label{A1}
Q_{\mu\nu\mu'\nu'}^{TT}(X,X') & = & {\cal{S}}{\cal{S}}' \Big[ F_1g_{\mu\nu}g'_{\mu'\nu'} + F_2(g_{\mu\nu}n_{\mu'}n_{\nu'} + g'_{\mu'\nu'}n_\mu n_\nu) + F_3n_\mu n_\nu n_{\mu'} n_{\nu'} + F_4g_{\mu\mu'}n_\nu n_{\nu'} + F_5g_{\mu\mu'}g'_{\nu\nu'} \Big], \nn\\
\end{eqnarray}
in which

$$F_1 = -\frac{8}{27}\frac{{\cal{Z}}(2 - 3{\cal{Z}}^2)}{(1 - {\cal{Z}}^2)^3},\;\;\;F_2 = -\frac{8}{27}\frac{{\cal{Z}}(-3 - 3{\cal{Z}}^2)}{(1 - {\cal{Z}}^2)^3},$$
$$F_3 = -\frac{8}{27}\frac{(-40 -114{\cal{Z}} -88{\cal{Z}}^2 + 12{\cal{Z}}^3 +32{\cal{Z}}^4 + 6{\cal{Z}}^5)}{(1 + {\cal{Z}})^2(1 - {\cal{Z}}^2)^3},$$
$$F_4 = -\frac{8}{27}\frac{(40 + 22{\cal{Z}} - 20{\cal{Z}}^2 - 18{\cal{Z}}^3)}{(1 - {\cal{Z}}^2)^3},\;\;\;F_5 = -\frac{8}{27}\frac{{\cal{Z}}(-11 + 9{\cal{Z}}^2)}{(1 - {\cal{Z}}^2)^3}.$$
Such a comparison can now be easily done by considering the tensor/vector sector of the graviton Wightman two-point function given in Ref. \cite{AHiguchi124006}; in the gauge $a = \alpha = 5/3$ which as already pointed out is employed to remove logarithmic divergences, we calculate the corresponding commutator in 4-dimension using Eqs. (A10a) to (A10e) in Ref. \cite{AHiguchi124006}. On this basis, the obtained commutator would be of the same structure as (\ref{A1}) (see Eqs. (172) and (173) in Ref. \cite{AHiguchi124006}) with the following coefficients:

$$F'_1 = -\frac{32}{27}\frac{{\cal{Z}}(1 + 3{\cal{Z}}^2)}{(1 - {\cal{Z}}^2)^3},\;\;\;F'_2 = \frac{32}{9}\frac{{\cal{Z}}(7 + {\cal{Z}}^2)}{(1 - {\cal{Z}}^2)^3},$$
$$F'_3 = -\frac{8}{27}\frac{{\cal{Z}}(985 - 263{\cal{Z}}^2 + 19{\cal{Z}}^4 + 27{\cal{Z}}^6)}{(1 - {\cal{Z}}^2)^3},$$
$$F'_4 = -\frac{2}{27}\frac{{\cal{Z}}(649 - 311{\cal{Z}}^2 + 19{\cal{Z}}^4 + 27{\cal{Z}}^6)}{(1 - {\cal{Z}}^2)^3},\;\;\;F'_5 = \frac{16}{27}\frac{{\cal{Z}}(-17 + 9{\cal{Z}}^2)}{(1 - {\cal{Z}}^2)^3}.$$

Both coefficients (`$F$'s) and (`$F'$'s) have the same overall structure but disagree in numerical factors which is due to the fact that the transverse-traceless and scalar sectors are defined differently in our paper and Ref. \cite{AHiguchi124006}. Moreover, contrary to our calculated Krein two-point function, the corresponding commutator obtained from \cite{AHiguchi124006} suffers from IR divergences. It is indeed because of different vacuums utilized in these works which has been thoroughly discussed in the present article.

\end{widetext}

\end{appendix}

\end{document}